\documentclass[]{interact}

\usepackage[version=3]{mhchem}
\usepackage{xcolor}
\usepackage{epstopdf}
\usepackage[caption=false]{subfig}

\usepackage[numbers,sort&compress,merge]{natbib}
\bibpunct[, ]{[}{]}{,}{n}{,}{,}
\renewcommand\bibfont{\fontsize{10}{12}\selectfont}

\theoremstyle{plain}

\theoremstyle{definition}

\theoremstyle{remark}

\begin{document}

\articletype{This is an original manuscript of an article published on line by Taylor and Francis in Molecular Physics on 22 Jun 2023, available at: https://doi.org/10.1080/00268976.2023.2223079}


\title{Infrared action spectroscopy as tool for probing gas-phase dynamics: Protonated Dimethyl Ether, \ce{(CH3)2OH+}, formed by the reaction of \ce{CH3OH2+} with \ce{CH3OH}}

\author{
\name{V. Richardson\textsuperscript{a, b, 1}, D.B. Rap\textsuperscript{c, 1}, S. Br\"{u}nken\textsuperscript{c} and D. Ascenzi\textsuperscript{b}\thanks{CONTACT D. Ascenzi. Email: daniela.ascenzi@unitn.it}}
\affil{
\textsuperscript{a}{Department of Physics, The Oliver Lodge, University of Liverpool, Oxford St, Liverpool, L69 7ZE, UK}\\
\textsuperscript{b}{Department of Physics, University of Trento, Via Sommarive 14, I-38123, Trento, Italy}\\
\textsuperscript{c}{Institute for Molecules and Materials, FELIX Laboratory, Radboud University, Toernooiveld 7, 6525 ED Nijmegen, The Netherlands}
}
}

\maketitle

\begin{abstract}
Methanol is one of the most abundant interstellar Complex Organic Molecules (iCOMs) and it represents a major building block for the synthesis of increasingly complex oxygen-containing molecules. The reaction between protonated methanol and its neutral counterpart, giving protonated dimethyl ether, \ce{(CH3)2OH+}, along with the ejection of a water molecule, has been proposed as a key reaction in the synthesis of dimethyl ether in space.

Here, gas phase vibrational spectra of the \ce{(CH3)2OH+}  reaction product and of the \ce{[C2H9O2]+} intermediate complex(es), formed under different pressure and temperature conditions, are presented. The widely tunable free electron laser for infrared experiments, FELIX, was employed to record their vibrational fingerprint spectra using different types of infrared action spectroscopy in the $600-1700$ cm$^{-1}$ frequency range, complemented with measurements using an OPO/OPA system to cover the \ce{O-H} stretching region $3400-3700$ cm$^{-1}$. The formation of protonated dimethyl ether as a product of the reaction is spectroscopically confirmed, providing the first gas-phase vibrational spectrum of this potentially relevant astrochemical ion.  

\end{abstract}

\begin{keywords}
Infrared action spectroscopy, ion-molecule reactions, cold ion spectroscopy, interstellar complex organic molecules (iCOMs), astrochemistry
\end{keywords}

\footnote{\textit{These authors contributed equally to the manuscript}}

\section{Introduction}


Starting with its first interstellar detection in 1970 \cite{Ball1970}, methanol (\ce{CH3OH}) has been detected in a wide range of astronomical environments that span the stages of stellar evolution from cold and dark pre-stellar cores to hot cores and corinos, as well as the gas-phase envelopes around young stellar objects of various masses and temperatures \cite{Spezzano2021, Lee2017, Belloche2020, PEACHES2021, Chahine2022, Mercimek2022}. Importantly, methanol has also been detected both in protoplanetary disks \cite{Walsh2016, vanthoff2018, Podio2020, vandermarel2021, Booth2021}, as a volatile species in a number of comets, including 67P/Churyumov-Gerasimenko which was recently studied as part of the Rosetta mission \cite{Leroy2015, Rubin2019, Altwegg2019}, and on the surface of a Kuiper Belt object \cite{Grundy2020}. This means that not only is methanol potentially present at the point of planet formation, but also in a wide range of bodies relevant to planetary astrochemistry. 

This is of particular significance as the reactivity of methanol presents a major route to the synthesis of increasingly complex organic species such as aldehydes, ketones and esters which, in turn, can then be precursors  to form prebiotic species such as amino acids, sugars and other biomolecules \cite{Sandford2020, Garrod2020, Oberg2021}. Understanding the reactivity of methanol and methanol derivatives is therefore of paramount importance to our  understanding of the evolution of oxygen-containing interstellar complex organic molecules (iCOMs), both in interstellar environments and planetary atmospheres. 

Dimethyl ether (DME, \ce{CH3OCH3}) is another highly-abundant O-containing iCOM. Following its first detection towards the Orion molecular cloud \cite{Snyder74}, DME has been found to be highly abundant in the warm, dense regions around numerous high-mass and low-mass protostars (the so-called hot cores and corinos) \cite{Nummelin2000, Bottinelli2007, Cazaux2003, Jorgensen2005, Favre2011, Bisschop2013,  Bianchi2019} and in the shocked regions driven by hot corinos \cite{SOLISX2020}. There, the $^{13}$C isotopologues \cite{Koerber2013} as well as the mono- and doubly-deuterated variants have been detected \cite{Richard2013, Richard2021}. However, DME is also present in objects exhibiting  different physical conditions, such as the colder and more rarefied environments of dark clouds and pre-stellar cores \cite{Bacmann2012, Cernicharo2012, Vastel2014, Jimenez, Agundez}, as well as the molecular clouds towards the Galactic centre \cite{Requena2006}. DME is also one of the largest molecules so far detected in other galaxies \cite{Sewilo2018}.

Despite being a ubiquitous and abundant molecule at different phases of star formation, the mechanisms leading to production of DME are still a matter of debate. A number of different formation routes have been reported in the literature, either via ice/grain chemistry or gas-phase chemistry (involving the reactions of both neutral and charged species) or a combination of both gas-phase and surface chemistry \cite{Charnley, Rodgers, Peeters, Brouillet2013, Balucani, Skouteris2019, Garrod2020a, Herbst2021, Tennis2021, Garrod2022}. 

Formation via surface chemistry, $e.g.$ via recombination of methyl (\ce{CH3}) and methoxy (\ce{CH3O}) radicals on dust-grain particles, is a feasible mechanism to explain DME abundances in the gas phase, but only when the dust grains are heated to sufficiently high temperatures to increase radical mobility and desorption from the grain surface \cite{Brouillet2013, Garrod2022}. Such mechanisms are not operative under the low-temperature conditions ($\sim 10$ K) of dark clouds and prestellar cores, and the warm case scenario has been challenged by the detection of DME (as well as other iCOMs) in these low temperature regions \cite{Bacmann2012, Cernicharo2012, Vastel2014, Agundez}. Various modifications to the surface models have been proposed, such as including reactive chemical desorption processes, introducing non-diffusive reactions between newly formed radicals and nearby species on the grains, adding alternative surface-formation processes ($e.g.$ \ce{H + CH3OCH2} or \ce{CH2 + CH3OH}) or introducing non-thermal chemistry triggered by cosmic rays \cite{Garrod2020a, Herbst2021}, but a consensus on the mechanisms operative in cold sources has not been reached as yet. 

With regards to the gas phase mechanisms, a recent theoretical study has investigated the formation via radiative association of \ce{CH3} and \ce{CH3O} radicals which, while found to proceed rapidly, still cannot fully explain observed DME abundances in cold and prestellar cores \cite{Tennis2021, Balucani}. Given this, another reaction of interest is the ion-molecule reaction corresponding to the self-methylation of methanol which is expected to produce protonated DME \cite{Charnley, Rodgers, Peeters, Brouillet2013}: 
\begin{align}
	\ce{ CH3OH + CH3OH2+ & \rightarrow 
 [C2H9O2]+ \rightarrow (CH3)2OH+ + H2O
	} 		
\end{align}
with the fate of the protonated molecule differing according to the environment. If dissociative recombination with electrons is the prevailing route, this would lead to the outright destruction of DME \cite{Hamberg2010}. But in the presence of alternative processes, the \ce{(CH3)2OH+} ion easily formed via reaction (1) can be efficiently converted into its neutral counterpart. As originally proposed in \cite{Taquet}, one such process is proton transfer to species with relatively large proton affinities such as \ce{NH3} \cite{Skouteris2019}:
\begin{align}
	\ce{ (CH3)2OH+ + NH3 & \rightarrow CH3OCH3 + NH4+
	} 		
\end{align}
In hot cores, following thermal desorption of large amounts of solid phase ammonia, the gas phase production via reaction (1) and (2) can be an efficient formation mechanism for DME \cite{Garrod2022}.

Reaction (1), that is thought to proceed via an intermediate proton-bound dimer (\textit{m/z} 65), has been studied extensively through both experimental kinetics and computational energetics studies \cite{Bass, Karpas, Morris, Dang, Bouchoux, Fridgen2001, White, Skouteris2019}, with rates reported in both KIDA \cite{KIDA2015} and UMIST \cite{UMIST2013}. However, as with most mass-spectroscopic gas-phase studies, identification of the \textit{m/z} 47 product of reaction (1) as protonated DME (\ce{(CH3)2OH+}), as opposed to its isomer protonated ethanol (\ce{CH3CH2OH2+}), has been made on the basis of computational work on the reaction pathways, as the majority of experimental techniques do not provide structural information on the products beyond their \textit{m/z} values. 

Furthermore, beyond the scope of the reaction studied as part of this work, as detection capabilities improve to the point where column densities and abundance measurements can be performed on an isomer-specific basis, as is the case with the Atacama Large Interferometer Array (ALMA) \cite{Cordiner19}, the ability to provide isomer-specific product branching ratios and reaction cross sections is likely to become essential for the development of models that can accurately replicate observations. 

One powerful method for extracting structural information of gas-phase molecular ions is infrared spectroscopy. Infrared Predissociation (IR-PD) spectroscopy, which makes use of cryogenic multipole radio-frequency (RF) ion traps and buffer-gas cooling, is a well-suited method to structurally investigate both stable product ions and weakly-bound ionic complexes \cite{Roithova, Schlemmer, Brummer, Gunther, Jusko, Schwartz, DeVine, Roithova2020}. At higher temperatures, and for larger ions, Infrared Multi-Photon Dissociation (IR-MPD) spectroscopy with powerful free-electron lasers is a versatile alternative to obtain structural information of molecular ions \cite{LeMaire2002, Valle2005, Oomens2006}, and can even be used for \textit{in situ} probing of reaction intermediates and products \cite{Polfer2007, Rap}. Both methods are therefore ideal tools for studying the isomer-specific kinetics of gas-phase ion-neutral reactions.

Whereas the rotational and ro-vibrational spectra of neutral DME have been extensively studied both experimentally and theoretically \cite{Senent1995a, Groner1998, Endres2009, Kutzer2016, Villa2011}, there is no spectroscopic information on its protonated variant and no experimental IR spectrum has been recorded to date, as earlier studies only targeted its proton-bound dimer \cite{Fridgen2001a, Fridgen2005}. The broadband vibrational fingerprint spectrum of its isomer, protonated ethanol, on the other hand, has recently been studied by IR-PD and IR-MPD action spectroscopy with the same cryogenic ion trap tandem-mass spectrometer coupled to the intense and widely tuneable free-electron lasers at the FELIX Laboratory as used in the present study \cite{Jusko}, and by IR-PD in the OH stretching fingerprint region \cite{Solca2005}. Though an IR-MPD spectrum of the \textit{m/z} 65 proton-bound dimer, the possible  intermediate in reaction (1), has been recorded previously \cite{Fridgen}, this has been achieved by producing the dimers by bimolecular ion–molecule displacement reactions involving \ce{(CHF2)2O} and methanol, and not in a direct association of protonated methanol with its neutral counterpart \cite{Fridgen2005}.
		
In this work we present IR-PD spectra for the \textit{m/z} 47 product ion of reaction (1), as well as IR-MPD spectra for \textit{m/z} 65 intermediate complexes formed both in an above-room-temperature Storage Ion Source (SIS) and in the cryogenic ion trap maintained at 200 K, where they can be stabilised by three-body collisions. In this way, we provide conclusive characterisation of the structure of the \ce{[C2H7O]+} ion produced as being protonated DME, as well as elucidating the structures of the stabilized intermediates and their formation mechanisms. 

\section{Methods}

\subsection{Experimental Details}

Experiments were performed using the FELion cryogenic ion trap end station at the Free Electron Lasers for Infrared eXperiments (FELIX) Laboratory \cite{Oepts}. The apparatus and its use in vibrational studies of molecular ions using both IR-PD and IR-MPD spectroscopy have been described in detail previously \cite{Jusko}, and its application for the elucidation of ion-molecule reaction pathways via \textit{in-situ} spectroscopic probing has been recently demonstrated \cite{Rap}. Here, we describe merely the details specific to the current investigation on the ions formed from the reaction of protonated methanol (\ce{CH3OH2+}) with methanol (\ce{CH3OH}). 

\ce{CH3OH2+} (\textit{m/z} 33), \ce{[C2H7O]+} (\textit{m/z} 47) and \ce{[C2H9O2]+} (\textit{m/z} 65) ions were produced from an evaporated sample of pure methanol (Sigma Aldrich, $>99.9$~\%) at a pressure of $\sim 10^{-5}$ mbar via electron impact ionization, at an energy of $\sim 15$~eV, and subsequent reactions in an ion storage source (SIS) at typical temperatures of $\sim 400$~K \cite{Gerlich}. Ions were extracted from the source in pulses of the order of a few tens of ms and selected by their mass-over-charge ratio (\textit{m/z}) using a quadrupole mass filter, before being guided into a 22-pole ion trap which is maintained at a fixed temperature in the $(5-200)$ K range. 

Here, for IR-PD spectroscopic measurements, ions are cooled to the trap temperature ($\sim$6 K in this case) through collision with a $3:1$ mixture of \ce{He}:\ce{Ne}, which is inserted into the trap via a pulsed piezo valve at a number density of $\sim 10^{14}$ cm$^{-3}$. Each pulse is triggered $10-15$ ms before the ions enter the trap and lasts for the length of the ion pulse, which allows for up to $\sim 60$\% of the primary ions to form weakly-bound complexes with \ce{Ne}. 
A representative mass spectrum of the tagging at \textit{m/z} 47 is given in Fig. S1 of the ESI, where we note the presence of a \ce{H2} impurity in the \ce{He}:\ce{Ne} mix which, while it leads to complexes of the ion with \ce{H2} and reduces the proportion of \ce{Ne}-tagged species present, is not expected to have a significant effect on the spectroscopy of the \ce{Ne}-tagged complexes.

IR-PD spectra are recorded by measuring the depletion $D=1-\frac{N(\nu)}{N_0}$ of the \ce{Ne}-tagged ion mass as a function of the laser frequency $\nu$, as the absorption of a resonant IR photon leads to dissociation of the complex ions. In order to account for potential saturation and baseline variability, as well as for differing laser pulse energies and pulse numbers, we normalise each spectrum to give intensity, $I$, in units of relative cross section per photon.
\begin{align}
	I = \frac{\ln(N(\nu)/N_0)}{n\cdot E/(h\cdot \nu)}
\end{align}
Where $N(\nu)$ is the number of complex ions as a function of the wavelength, $N_0$ is the baseline ion count, $n$ is the number of pulses and $E$ is the laser pulse energy.

IR-MPD spectra are recorded by measuring the depletion of the bare ion counts as a function of the wavelength, with the absorption of multiple resonant photons leading to a vibrational pumping of the ion that eventually results in fragmentation along the lowest energy pathway.
The IR-MPD method was used to spectroscopically probe intermediate and product ions produced either in the source or in the trap from the reaction of \ce{CH3OH2+} with \ce{CH3OH}. Here, the trap temperature is held at higher values (in the range of 193-208~K for the reaction in the trap and 5-29~K for the reaction in the SIS), and pure He is used for thermaliation and trapping to avoid forming ion rare-gas complexes. 
For reactions in the trap, methanol is introduced in a highly diluted mix
with \ce{He} as a $\sim 10 $ ms pulse reaching peak pressures of $\sim 1 \cdot 10^{-3}$ mbar inside the trap at the beginning of the trapping cycle. Depletion spectra of the \textit{m/z} 65 ion formed both in the trap and the SIS have been measured with a typical trapping time of 0.6 s to avoid saturation of the strong bands, while a longer time of 1.6 s was used for gain spectra of the \textit{m/z} 33 and 47 mass channels. The intensities for the IR-MPD spectra are given either as relative depletion in percentage or as intensities normalized for the laser power. 

IR radiation was provided by the FEL-2 free-electron laser of the FELIX Laboratory \cite{Oepts} operated at 10 Hz in the $600-1700$ cm$^{-1}$ wavenumber region with macropulse energies inside the ion trap of up to 11 mJ, and a FWHM of approximately 1$\%$ of the laser frequency. The \ce{O-H} stretching region between $3400-3700$ cm$^{-1}$ was scanned using a tabletop Laservision OPO/OPA system, again operated at 10 Hz, with a FWHM of $\sim$1 cm$^{-1}$, a 5 ns pulse length and energies of $\sim10$ mJ.

Saturation depletion measurements have also been performed in order to quantify isomeric abundances and purities for a given \textit{m/z}, as previously described in details \cite{Jusko, Marimuthu}. In brief, this involves the exposure of the \ce{Ne}-tagged ions to a series of laser pulses, resonant with an isomer-specific vibrational band, in order to fully dissociate all \ce{Ne}-tagged complexes of this isomer. The fractional abundance of this isomer (hereafter referred to simply as the $A$ parameter) can then be determined by fitting the depletion of ion counts as a function of the deposited energy: $D=A\cdot (1-e^{-K_{on}\cdot n\cdot E})$, with $K_{on}$ the rate coefficient of dissociation on resonance. 

\subsection{Quantum Chemical Calculations}

The experimental IR spectra have been interpreted through comparison with vibrational frequencies from density functional theory (DFT) calculations using Gaussian 16 \cite{Frish}. The structures and harmonic vibrational frequencies of protonated methanol, protonated ethanol and protonated DME were calculated using the double hybrid B2PLYP functional with the D3 Grimme dispersion correction method, and aug-cc-pVTZ basis set combination \cite{Frish, Schwabe, Barone2015}. Typical frequency scaling factors of 0.976 and 0.956 were used for the fingerprint and OH stretching region, respectively \cite{Bauschlicher_2018}. Comparison of the predicted C-O-C stretching vibration of protonated DME using this method with the predictions from harmonic and anharmonic calculations at the B3LYP-GD3/N07D level of theory is given in Fig. S2 in the ESI. 

Five different \ce{[C2H9O2]+} isomers have been identified by previous studies as potentially contributing to the reaction pathway \cite{Fridgen2001, Skouteris2019, White}, and geometry optimization and harmonic frequency calculations have been performed for each of these structures. After a benchmarking of the different function/basis set combinations performed on some of the isomers via comparison with the observed experimental spectrum, the calculations shown in this work are those performed using M06-2X functional \cite{Zhao} with the 6-311++G(d,p) basis set.

Furthermore, transition states between the two conformers, \textit{cis} and \textit{trans} proton-bound methanol dimer structures, not previously discussed in the literature, has been identified as part of this work. Transition state geometries have been optimized at the M062X/6-311++G(d,p) level of theory and single-point energy calculations of both conformers and the connecting transition states have been performed at the MP2/6-311++G(2d,2p)//M062X/6-311++G(d,p) level of theory.

\section{Results and Discussion}

\begin{figure}
    \centering
    \includegraphics[width=13cm]{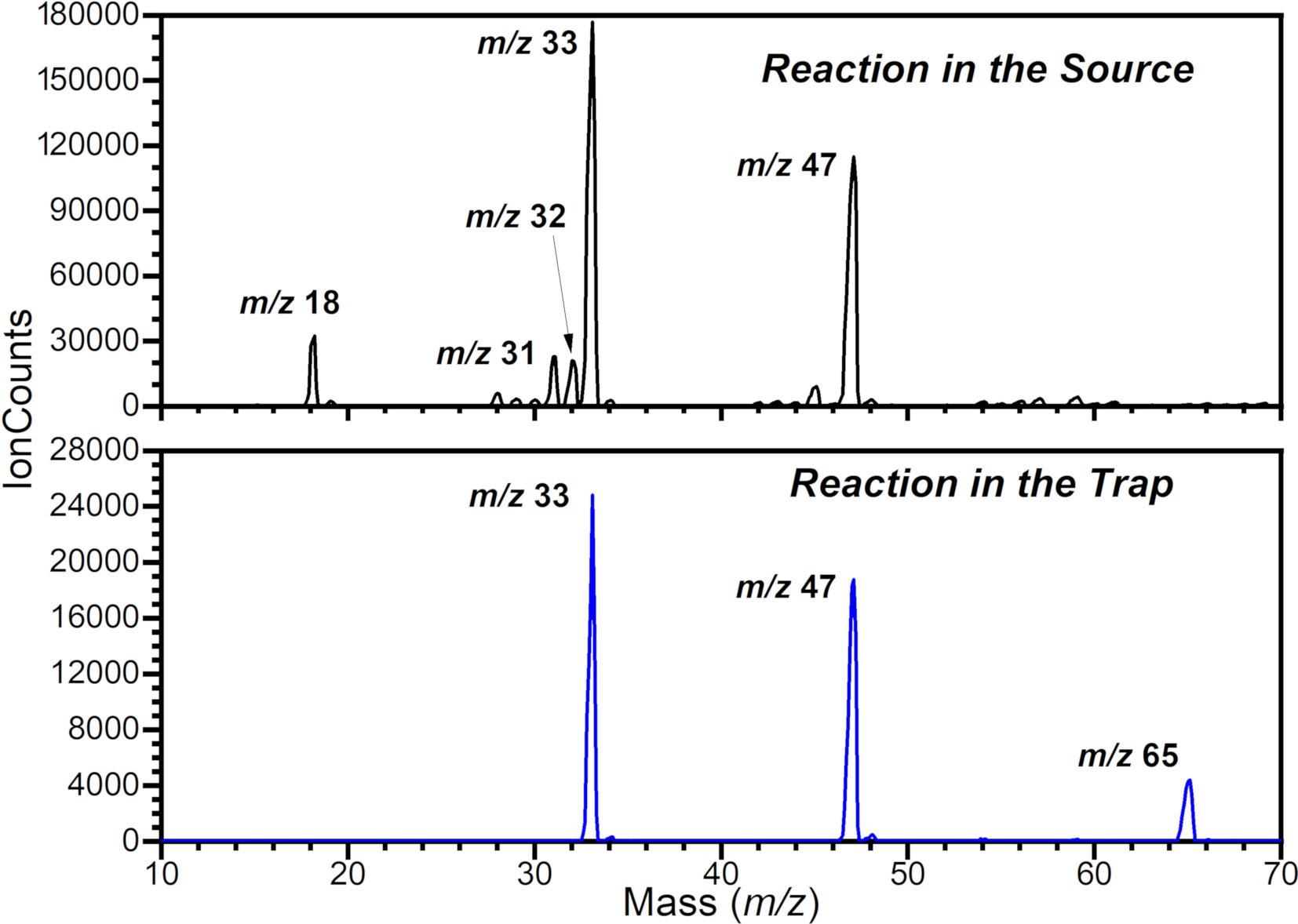}
    \caption{Mass spectra of the products of the reaction of \ce{CH3OH2+} (\textit{m/z} 33) with \ce{CH3OH} in both the SIS (top) and trap (bottom).}
    \label{fig:MSspectra}
\end{figure}

In this work, reaction (1) has been studied both in the SIS and in the trap, with representative mass spectra of the reaction products shown in Figure \ref{fig:MSspectra}. In both cases, we observe the \textit{m/z} 33 reactant ion (\ce{CH3OH2+}) and \textit{m/z} 47 product ion (\ce{[C2H7O]+}), with the mass spectrum for the reaction in the trap also exhibiting a minor peak at \textit{m/z} 65 corresponding to one or more \ce{[C2H9O2]+} complexes, which are efficiently stabilised due to the 100-fold higher pressure in the trap than the SIS.

For the reaction in the SIS, we also note the presence of a series of less intense peaks at \textit{m/z} 18 (\ce{H2O+}), 31 (\ce{CH2OH+}), 32 (\ce{CH3OH+}) and 45 (\ce{[C2H5O]+}). The \textit{m/z} 18 and 32 products are the result of non-dissociative ionization of background gas containing water and reactant methanol (or background \ce{O2}) respectively, while the \textit{m/z} 31 and 45 ions are the result of \ce{H2}-ejection from the \textit{m/z} 33 reactant ion and \textit{m/z} 47 product ion respectively. By increasing the pressure in the source to $\sim 3.5$ x$10^{-5}$~mbar we are also able to stabilize the \textit{m/z} 65 association complexes through three-body collisions, thereby allowing for a spectral comparison of both product and intermediate ions generated in the SIS and in the trap.

In order to check the isomeric purity of the \textit{m/z} 33 precursor ion, a Ne-tagging IR-PD spectrum in the O-H stretching region has been obtained using a table-top OPO/OPA system, with the results given by Fig. S3 in the ESI. Two intense bands at 3492(2) and 3563(2) cm$^{-1}$ were observed, in agreement (to within $\sim$20 cm$^{-1}$) with calculated scaled harmonic frequencies of the symmetric and antisymmetric \ce{O}-{H} stretching modes of protonated methanol (\ce{CH3OH2+}). Furthermore, the observed band positions agree to within $\sim$10 cm$^{-1}$ with earlier measurements using He-tagging IR-PD and Laser Induced Inhibition of Complex Growth (LIICG) \cite{Jusko}, and the observed red-shift can be rationalized by the fact that tagging with \ce{Ne} atoms, as performed here, leads to more strongly-bound complexes than equivalent tagging with \ce{He}. Saturation depletion measurements on these bands, as described in Section 2.1, revealed an isomeric purity of 98(5) \%, see Fig. S4 in the ESI. From this, we are therefore able to conclude that we are selectively generating the desired \ce{CH3OH2+} reactant ion, protonated methanol, at \textit{m/z} 33.

Though the formation of the \ce{CH3CH2OH2+} (protonated ethanol) isomer of \ce{[C2H7O]+} is $\sim$35 kJ/mol lower in energy than the equivalent channel leading to protonated DME, \ce{(CH3)2OH+}, \cite{ATcT, Hunter}, early studies tentatively assigned the \textit{m/z} 47 product ion to the formation of the latter species \cite{Munson}. This conclusion has since been reinforced by a collision-induced dissociation study \cite{Graul} as well as a number of computational studies on the reaction Potential Energy Surface (PES) \cite{Fridgen, Bouchoux, Skouteris2019}. However, to the best of our knowledge, no spectroscopic studies of the reaction product has so far been performed. 

The first computational work on the PES for the \ce{CH3OH2+} + \ce{CH3OH} system  \cite{Bouchoux} identified five \textit{m/z} 65 isomeric complexes as intermediates, namely the proton-bound dimer of methanol (in both \textit{cis} and \textit{trans} conformers), the \ce{CH3}-bound protonated methanol dimer, the \ce{H2O}-release complex and the proton-bound complex of the products (DME and water).
Though subsequent works on the PES \cite{Fridgen2001, Skouteris2019, White} have reported slight changes in the energetics of the different structures, no additional complexes are considered. For this reason, here we have limited consideration of the various contributions to the \textit{m/z} 65 IR-MPD spectra to these five structures, whose geometries, energies and interconversion pathways will be discussed in details in Sec. \ref{sec:m65}. 

\subsection{Structural Characterization of the \textit{m/z} 47 Product Ion}

\begin{figure}
   \centering
   \includegraphics[width=14cm]{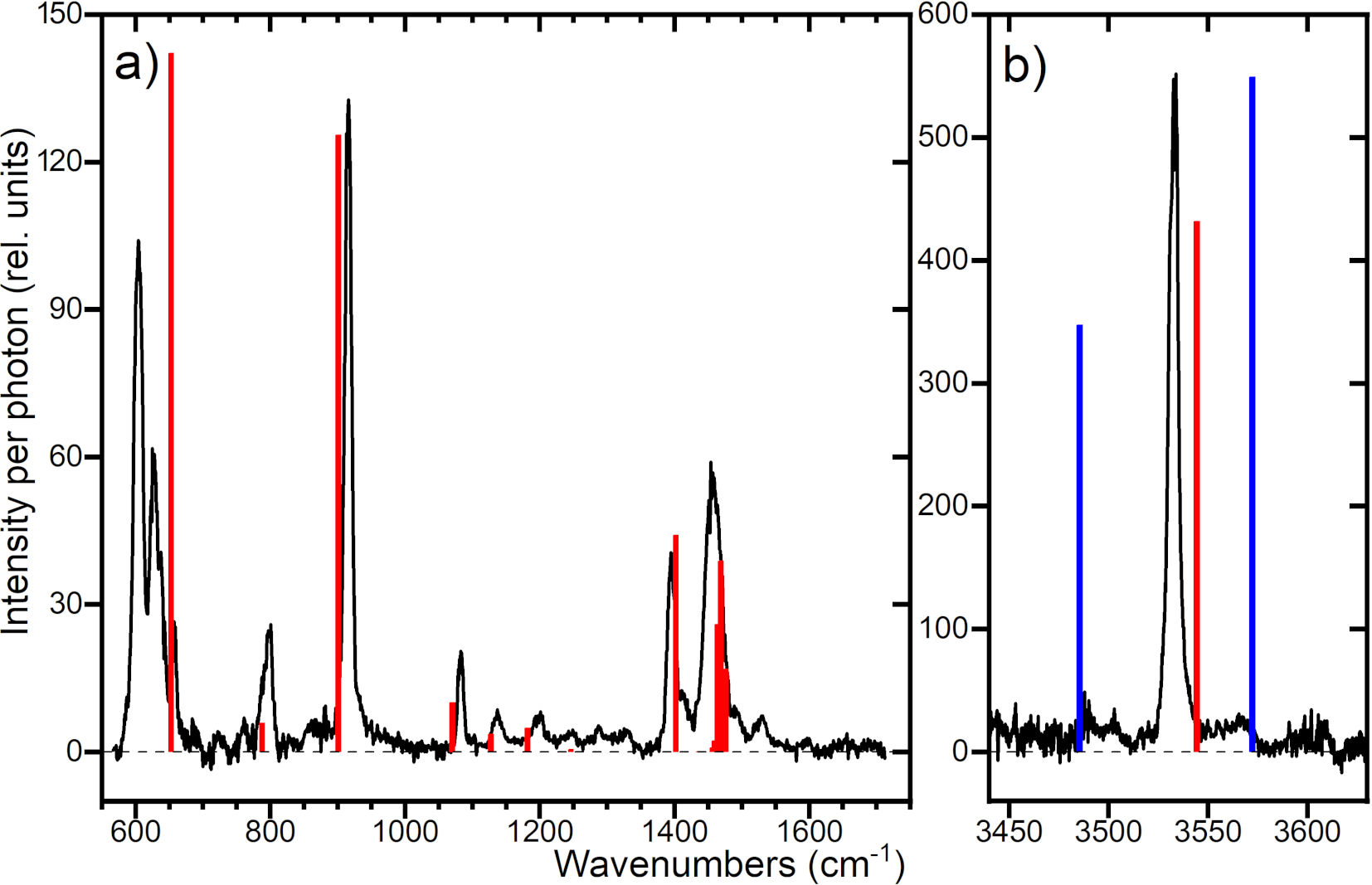}
   \caption{Comparison of the IR-PD spectrum of the \ce{Ne}-tagged \textit{m/z} 47 ion generated by the reaction of \ce{CH3OH2+} with \ce{CH3OH} in the FELion SIS measured using (a) the FELIX light source and (b) a tabletop OPO laser (both black lines) with the harmonic computational spectra for the \ce{(CH3)2OH+} (red bars) and \ce{CH3CH2OH2+} ions (blue bars, not shown in the fingerprint region for clarity reasons) at the B2PLYPD3/aug-cc-pVTZ level of theory. A frequency scaling factor of 0.976 and 0.956 was applied to the calculated bands in the fingerprint and \ce{O-H} stretching region, respectively.}
   \label{fig:m47spectrum}
\end{figure}

In order to characterize the \textit{m/z} 47 ion formed by reaction (1) in the SIS we have collected IR-PD spectra for both the $500-1500$~cm$^{-1}$ and $3400-3700$~cm$^{-1}$ regions using FEL-2 and the OPO/OPA system, respectively, with the results shown in Fig. \ref{fig:m47spectrum} alongside the calculated IR spectra for protonated DME at the B2PLYPD3/aug-cc-pVTZ level of theory. In the \ce{O}-\ce{H} stretching region we also show the calculated band positions of the protonated ethanol isomer at the same level of theory. 

In the \ce{O-H} stretching region probed using the OPO laser, we observe a single intense band at 3532(2) cm$^{-1}$, while in the lower wavenumber region we observe a major intense band at 916(2) cm$^{-1}$ with weaker features at 603(2), 629(2), 657(2), 797(2), 1082(2), 1136(2), 1208(3), 1393(2), 1414(3) and 1458(2) cm$^{-1}$. An overview of the fitted band positions and their integrated relative intensities is given in Table \ref{tab:1} alongside equivalent computed values for the \ce{(CH3)2OH+} isomer at the B2PLYPD3/aug-cc-pVTZ level and subsequent assignments. 

\begin{table}
    \caption{IR-PD experimental vibrational band positions, FWHM (both in cm$^{-1}$) and integrated relative intensities of \ce{(CH3)2OH+} (absolute uncertainties in parenthesis) compared to calculated infrared frequencies (in cm$^{-1}$) and relative intensities (in km/mol) at the B2PLYPD3/aug-cc-pVTZ level of theory. Frequency scaling factors of 0.976 and 0.956 were applied to the calculated bands in the fingerprint and OH stretching region, respectively. 
    \label{tab:1}}
    \begin{tabular*}{\textwidth}{c|c|c|c|c|c}
        \hline
        Pos. & FWHM & Rel. Int. & Calc. Pos. & Calc. Rel. Int. & Mode\\
        \hline
         - & - & - & 148.0 & 0.1 & $\nu_{24}$ \\
         - & - & - & 200.8 & 0.9 & $\nu_{23}$ \\
         - & - & - & 328.7 & 6.5 & $\nu_{22}$ \\
        603 (2)  & 14 & 16.3 (0.6) & 652.6 & 142.1 & $\nu_{21}$ \\
        629 (2)  & 22 & 12.4 (0.5) & & & \\
        657 (2)  & 9  & 2.2 (0.2)  & & & \\
        797 (2)  & 16 & 3.8 (0.3)  & 787.7 & 5.8 & $\nu_{20}$ \\
        916 (2)  & 11 & 21.4 (0.6) & 900.9 & 125.5 & $\nu_{19}$ \\
        1082 (2) & 8  & 2.1 (0.1)  & 1069.9 & 10.1 & $\nu_{18}$ \\
        1136 (2) & 9  & 0.9 (0.1)  & 1127.4 & 3.6 & $\nu_{17}$ \\
        1208 (3) & 42 & 2.3 (0.2)  & 1181.8 & 4.9 & $\nu_{16}$ \\
        1393 (2) & 12 & 7.6 (0.4)  & 1401.9 & 44.1 & $\nu_{14}$ \\
        1414 (3) & 17 & 1.8 (0.3)  &        &      &            \\
        1458 (2) & 24 & 19.3 (0.7) & 1463.5/1468.8/1476.2 & 25.9/38.8/16.8 & $\nu_{10}$/$\nu_{9}$/$\nu_{8}$ \\
        3533 (2) & 6  & 10.2 (0.3) & 3544.3 & 215.8 & $\nu_{1}$\\
        \hline
    \end{tabular*}
\end{table}

As with the IR-PD spectrum of the \textit{m/z} 33 ion from the source, we note that the calculated \ce{O}-\ce{H} stretching frequency for \ce{(CH3)2OH+} is shifted with respect to that observed experimentally, this time to higher wavenumbers by $\sim$11 cm$^{-1}$. However, despite this shift, we note a good agreement with the computational spectrum, showing a single intense band in this region. We are able to exclude the presence of a significant amount of the \ce{CH3CH2OH2+} isomer by comparison to both its computational and previous experimental He-tagged IR-PD spectra \cite{Jusko} in the \ce{O}-\ce{H} stretching region, where a pair of bands corresponding to the symmetric and antisymmetric \ce{O-H} stretching modes, separated by $\sim$90 cm$^{-1}$, should be observed, which is inconsistent with the experimental data. These findings are supported by saturation depletion measurements described below.   

In the lower wavenumber region, the computed \ce{(CH3)2OH+} spectrum effectively reproduces the intense 916(2) cm$^{-1}$ band, corresponding to the $\nu_{19}$ \ce{C-O-C} asymmetric stretch, as well as the weaker features at 797(2), 1082(2), 1136(2), 1208(3), 1393(2) and 1458(2) cm$^{-1}$, with  consistent relative intensities, and the ability to predict the broadening of the 1458(2) cm$^{-1}$ structure due to the presence of multiple overlapping bands. The currently unassigned band observed at 1414(3) cm$^{-1}$ could be a combination or overtone band of the bending modes observed below 800 cm$^{-1}$. 

Whereas the scaled harmonic frequencies of the bare ion agree, to within 20 cm$^{-1}$, with those of the strong observed bands in this region, the region below 700 cm$^{-1}$ is not well reproduced. The experimental spectrum shows a triplet of bands significantly shifted with respect to the calculated frequency of the $\nu_{21}$ \ce{O-H} out of plane bending mode. We can only speculate that the additional bands are overtone or combination modes of the low-lying torsional modes $\nu_{23, 24}$ and the \ce{C-O-C} bending mode $\nu_{22}$, that might also interact with the $\nu_{21}$ bending mode. In neutral DME a strong interaction of the overtones of the two torsional modes with the \ce{C-O-C} bending mode is observed  \cite{Senent1995b,Villa2011}. Interestingly, the torsional and \ce{C-O-C} bending modes are predicted at lower frequencies than in neutral DME due to the protonation on the central \ce{O} atom. A theoretical treatment including anharmonic effects of these large-amplitude modes is beyond the scope of the present work.

IR-PD spectroscopy of the isomeric \ce{CH3CH2OH2+} ion using \ce{He}-tagging has been performed using this apparatus previously \cite{Jusko}, with intense bands observed at 617 and 665 cm$^{-1}$, and weaker bands observed at 949, 1389 and 1605 cm$^{-1}$. While both the computed and experimental spectra for the \ce{CH3CH2OH2+} ion show some agreement with the IR-PD spectrum observed here, they are unable to reproduce either the full range of bands or the relative intensities. Most notably, the intense band at $\sim$700 cm$^{-1}$, the \ce{OH2} wagging mode, is not observed experimentally, the 797(2) cm$^{-1}$ band in the experimental spectrum is not predicted, and, although having a slight overlap, the intensities of the 916(2) and 1458(2) cm$^{-1}$ bands observed in the experimental spectrum are not reproduced. On this basis, and considering the lack of agreement in the \ce{O}-\ce{H} stretching region, we are confident in assigning the experimental spectrum to the \ce{(CH3)2OH+} isomer, protonated DME. 

\begin{figure}
    \centering
    \includegraphics[width=14cm]{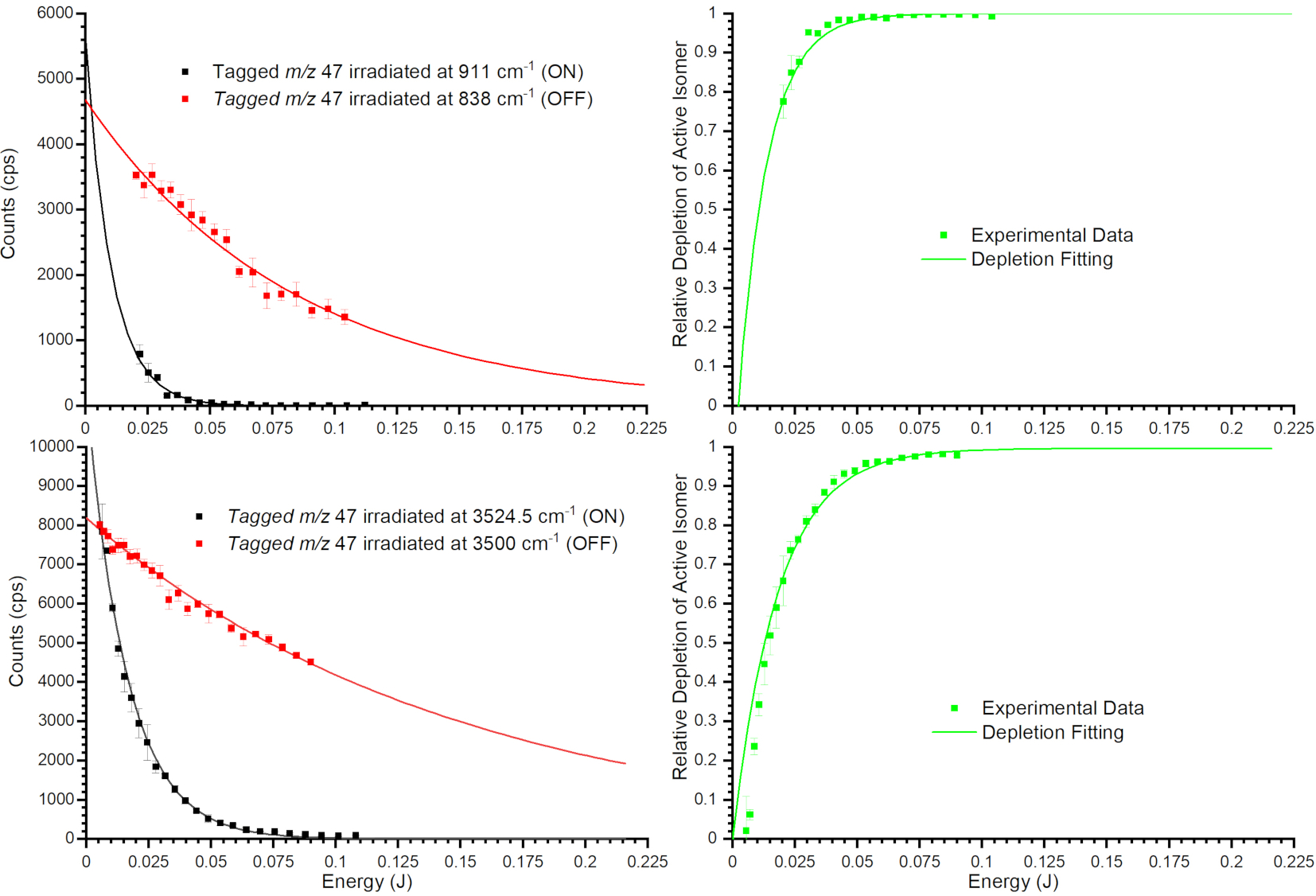}
    \caption{Depletion analysis of the \ce{(CH3)2OH+} ion. \textit{Top left:} On resonance depletion at 915 cm$^{-1}$ (black squares) compared with natural losses of ions off resonance (838 cm$^{-1}$, red squares) and fits thereof (black and red lines, respectively). \textit{Top right:} Relative depletion of active isomer as a function of deposited energy at  915 cm$^{-1}$, showing experimental data (green squares) and fit thereof (green line). \textit{Bottom left:} On resonance depletion at 3524.5 cm$^{-1}$ (black squares) compared with natural losses of ions off resonance (3500 cm$^{-1}$, red squares) and fits thereof (black and red lines, respectively). \textit{Bottom right:} Relative depletion of active isomer as a function of deposited energy at 3524.5 cm$^{-1}$, showing experimental data (green squares) and fit thereof (green line).}
	\label{fig:m47deplet}
\end{figure}

As with the \textit{m/z} 33 reagent ion, we have performed saturation depletion measurements, shown in Fig. \ref{fig:m47deplet}, for two unblended bands at 916(2) cm$^{-1}$ and 3533(2) cm$^{-1}$, from which we extracted values for the isomeric purity of 99(5) \% in both cases. From this, we are able to conclude that the \textit{m/z} 47 ion formed by reaction (1) in the SIS corresponds exclusively to \ce{(CH3)2OH+}.

Although we have also formed the \textit{m/z} 47 ion via the reaction in the trap at 200 K (see Fig. \ref{fig:MSspectra}), we were unable to perform subsequent IR-MPD measurements of this ion because the \textit{m/z} 65 complex ions, which are also formed and stabilised in the trap, fragment to the \textit{m/z} 47 mass channel upon resonant excitation, thereby preventing us from isolating the dissociation-induced depletion of the \textit{m/z} 47 ion. We have, however, for reference collected an IR-MPD spectrum of the \textit{m/z} 47 ion formed in the SIS, which exhibits a general agreement with  band positions of the IR-PD spectrum shown in Fig. \ref{fig:m47spectrum}, though with all bands exhibiting notable broadening and red-shifting. A comparison of the IR-PD and IR-MPD spectra for the \textit{m/z} 47 ion formed in the SIS is given in Fig. S5 in the ESI. 

\subsection{Structural Characterization of the \textit{m/z} 65 Intermediate}
\label{sec:m65}

In order to probe the underlying dynamics of reaction (1), we have also measured the infrared fingerprint of the \textit{m/z} 65 intermediate formed in both the trap and the SIS. However, in order to interpret these results, we must first consider the different \ce{[C2H9O2]+} isomers that are expected to contribute to the reaction PES and their various interconversions.  

Calculations of the \ce{CH3OH2+} plus \ce{CH3OH} reaction PES have been performed at different levels of theory, starting with Bouchoux and Choret \cite{Bouchoux} and then continued by Fridgen et al. \cite{Fridgen2001}, Skouteris et al. \cite{Skouteris2019} and White et al. \cite{White}. The transition state between the two proton-bound conformers has been identified previously by Bouchoux and Choret \cite{Bouchoux} as corresponding to the rotation of the protonated methanol in the \ce{H-O-O} plane, rather than as the barrier to interconversion through rotation around the \ce{O-H-O} bond axis. By performing a rigid potential energy scan for rotation around the \ce{O-H-O} bond axis, we have identified two transition states located at a much lower energy, each with the calculated frequencies exhibiting one imaginary frequency corresponding to the rotation around the \ce{O-H-O} bond axis. At the level of theory used, MP2/6-311++G(2d,2p)//M062X/6-311++G(d,p), the energy difference for the two conformers is within the error of the value calculated previously \cite{Bouchoux}, and so this earlier value is retained. For the rotation between the two conformers, energy barriers of $\sim$ 2 kJ/mol with respect to the \textit{trans} conformer are identified, with slight differences observed for rotation and counter-rotation due to the different sterical effects along the trajectories. Here, an averaged value of 2 kJ/mol is used, with full results shown graphically in Fig. S6 of the ESI.

\begin{figure}
	\centering
        \includegraphics[width=13cm]{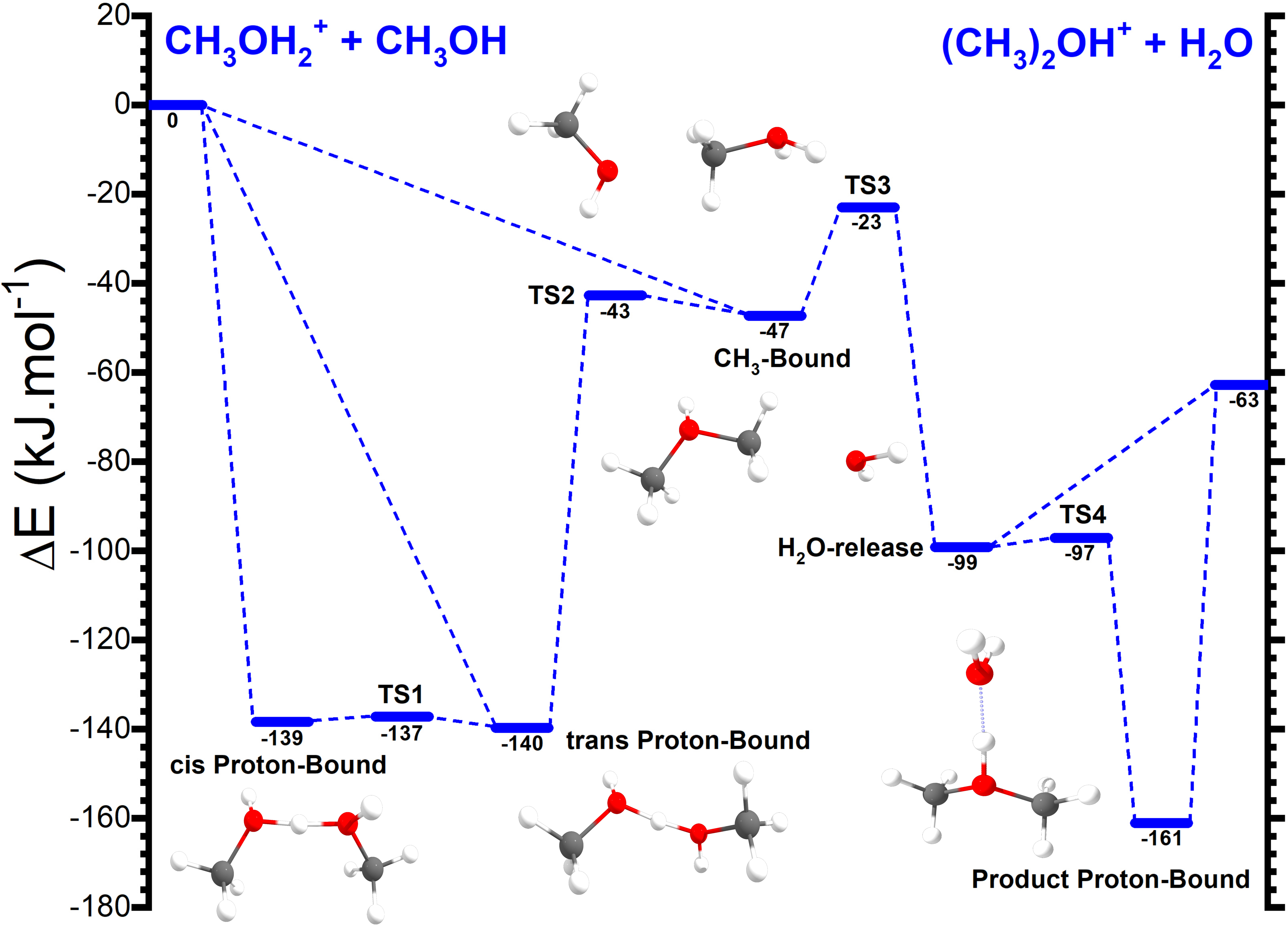}
	\caption{Schematic of the relevant minima and transitions states of the Potential Energy Surface for the reaction of \ce{CH3OH2+} with \ce{CH3OH}. Relative energies of the \textit{cis} and \textit{trans} proton-bound methanol dimers at the B3LYP/6-31+G\text{**} level of theory are taken from Fridgen et al. \cite{Fridgen}, with the barrier height for the \textit{cis-trans} isomerisation relative to the \textit{trans} isomer calculated at the MP2/6-311++G(2d,2p)//M062X/6-311++G(d,p) level of theory as part of this work. All other energies and connectivities are taken from Skouteris et al. \cite{Skouteris2019}, where calculations have been performed at the CCSD(T)/aug-cc-pVTZ level.}
    \label{fig:PES}
\end{figure}

A schematic with the relevant minima and transition states of the PES for reaction (1) is shown in Fig. \ref{fig:PES}, where the values calculated at the CCSD(T)/aug-cc-pVTZ level from Skouteris et al. \cite{Skouteris2019} have been complemented by the addition of relative energies of the \textit{cis} conformer of the \ce{CH3OH} proton bound dimer (calculated at the B3LYP/6-31+G\text{**} level by Fridgen et al. \cite{Fridgen2001}) and the \textit{cis-trans} isomerisation barrier calculated at the MP2/6-311++G(2d,2p)//M062X/6-311++G(d,p) level of theory as part of this work. 

Starting from the reactants, there is the possibility to form either conformer of the proton-bound dimer as well as the \ce{CH3}-bound dimer. However, a previous experimental study using \ce{^{18}O} and deuterium labelling to monitor the reaction through proton exchange, noted that the majority of the reactive flux is expected to pass, at least initially, through the two proton-bound dimers \cite{Dang}. 

Once formed, these two conformers can interconvert via the shallow transition state \textbf{TS1}, with the inter-conversion to the \ce{CH3}-bound dimer proceeding from the \textit{trans} conformer via \textbf{TS2}. From the \ce{CH3}-bound dimer, the reaction can proceed to the \ce{H2O}-release complex via a further, shallow, transition state, \textbf{TS3}. 

From the \ce{H2O}-release complex, there is the possibility to either separate barrierlessly into the products (observed at \textit{m/z} 47) or to undergo a further inter-conversion into the proton-bound complex of the products via \textbf{TS4}, with this final intermediate structure also able to separate into the products without a barrier. 

\begin{figure}
    \centering
    \includegraphics[width=13cm]{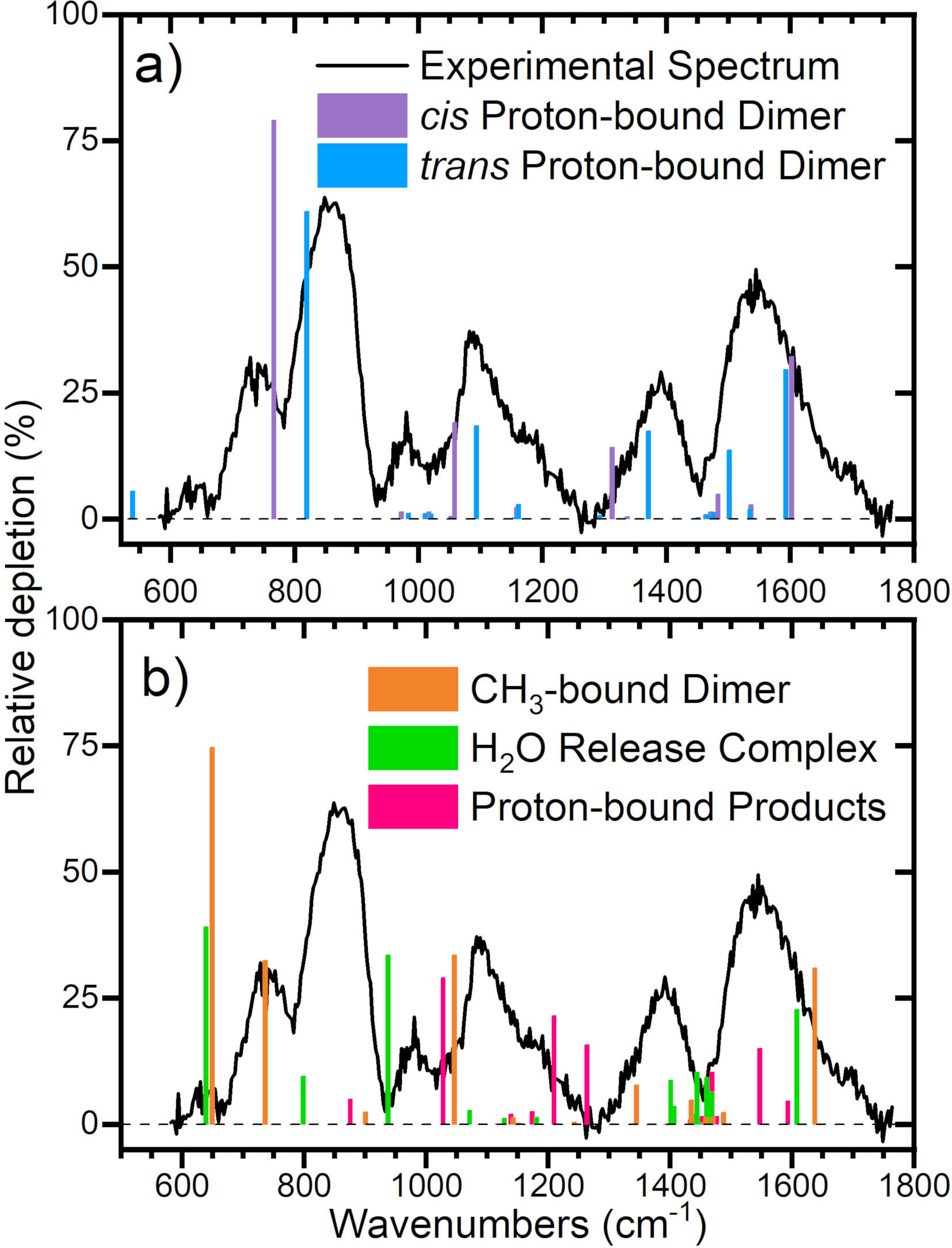}
    \caption{The \textit{m/z} 65 IR-MPD spectrum from the reaction of \ce{CH3OH2+} with \ce{CH3OH} in the FELion trap measured using the FELIX light source (black). Comparison with the computational spectra for the (a) 
    \textit{trans} and \textit{cis} Proton-Bound complexes (blue and purple, respectively) and (b) \ce{H2O}-release (green), Methyl-Bound dimer (orange) and the Product Proton-Bound (pink) complexes calculated at the M06-2X/6-311++G(d,p) level of theory. The calculated intensities of the proton-bound complexes have been scaled by a factor of 0.125 with respect to the other complexes.}
    \label{fig:m65IRMPD}
\end{figure}

The experimental IR-MPD spectrum of the complexes formed by the reaction in the trap, measured by monitoring the depletion of the \textit{m/z} 65 signal, is shown in Fig. \ref{fig:m65IRMPD}, alongside computed spectra of the five intermediates identified above. 

From Fig. \ref{fig:m65IRMPD}, we note an intense band at 851(2) cm$^{-1}$, intermediate intensity bands at 734(3), 1098(3), 1386(3) and 1553(3) cm$^{-1}$ and some potential weaker features at 624(3), 649(3) and 981(3) cm$^{-1}$. The intense band at 851(2) cm$^{-1}$ is in good agreement with the intense \ce{O-H-O} stretching vibration (a characteristic of proton-bound dimers) identified in the computed spectra for both conformers of the methanol proton-bound dimer. At the level of theory used in this work, these features have band positions of 766 and 819 cm$^{-1}$ for the \textit{cis} and \textit{trans} conformers respectively. Furthermore,  the bands at 1098(3), 1386(3) and 1553(3) cm$^{-1}$ are also consistent with features present in the computational spectra of both conformers. 

We have performed geometry optimizations and harmonic frequency calculations using four different functional/basis set combinations to assess if this difference also shows up when calculated using other methods (Fig. S7 in the ESI). The \ce{O-H-O} stretching vibrations of the \textit{cis} and \textit{trans} conformers are calculated at two significantly different frequencies when using the functionals M06-2X and $\omega$B97XD, due to differences in the \ce{O-H} bond lengths. The B3LYP-GD3 and DSD-PBEP86 functionals predict very similar \ce{O-H-O} stretching vibrations for both conformers. However, both $\omega$B97XD and especially M06-2X show a significantly better match with the experimental spectrum for the other vibrational modes. We conclude that the structures optimized using the latter functionals represent the actual equilibrium structure better, confirming the different \ce{O-H-O} stretching frequencies of the two proton-bound conformers. 

In order to accurately assign the band at 851(2) cm$^{-1}$, we must therefore also consider the source of the weaker feature at 734(3) cm$^{-1}$. Through comparison with the various computed spectra, one might assign the band to either the  799 cm$^{-1}$ band in the spectrum of the \ce{CH3}-bound complex or the 736 cm$^{-1}$ band of the proton-bound product complex. However, for both assignments we would expect to observe significantly more intense bands at 639 cm$^{-1}$ and 649 cm$^{-1}$ for the two conformers, respectively.

The most plausible assignment of the experimental 734(3) cm$^{-1}$ band is therefore the aforementioned 766 cm$^{-1}$ \ce{O-H-O} stretching band of the \textit{cis} proton-bound complex, thereby allowing for the assignment of the major 851(2) cm$^{-1}$ band to the calculated 819 cm$^{-1}$ \ce{O-H-O} stretching band of the \textit{trans} proton-bound complex. On this basis, other predicted bands of the \textit{trans} conformer are likely contributors to the observed depletion signal at 1098(3) and 1386(3), while the band at 1553(3) cm$^{-1}$ is tentatively assigned to a blend of the 1501 and 1593 cm$^{-1}$ bands in the computed spectrum of the \textit{trans}  proton-bound complex and the 1602 cm$^{-1}$ band of the \textit{cis} proton-bound complex, though we are unable to provide a more definite assignment.

Notably, the weaker features at 624(3) and 649(3)cm$^{-1}$ cannot be explained by the computed spectra for either of the proton-bound complexes. However, they could feasibly correspond to some combination of the \ce{H2O}-release and the 649 cm$^{-1}$ band in the spectrum of the \ce{CH3}-bound complex. Similarly, the other weak feature at 981(3) cm$^{-1}$ is also in reasonable agreement with either the 938 cm$^{-1}$ band in the computed \ce{H2O}-release spectrum, the 1028 cm$^{-1}$ band in the computed spectrum for the proton-bound complex of the products, or the 1046 cm$^{-1}$ band in the computed spectrum for the \ce{CH3}-bound dimer spectrum. 

We also cannot dismiss the possibility that the weaker unexplained features could be the result of anharmonic effects leading to overtone and combination bands that have not been treated using the performed harmonic frequency calculations. To this end, we note that the most intense lower-frequency bands are the vibrations (including frequency scaling of 0.976) at 323 and 389 cm$^{-1}$ of the \textit{cis} isomer and three features at 296, 406 and 539 cm$^{-1}$ of the \textit{trans} isomer, and that all, except for the 539 cm$^{-1}$ mode, can be attributed to the bending vibrations of the non-hydrogen-bonded OH groups. 

Previous IR-MPD spectroscopy studies of methanol proton-bound dimers have generated the complexes through the sequential reaction of \ce{CHF2+} with \ce{(CHF2)2} and two \ce{CH3OH} molecules \cite{Fridgen, Fridgen2005}, with bands observed at 866, 982, 1070-1225, 1330-1425 and 1467-1634 cm$^{-1}$, in good agreement with our study. Though they could not make a clear assignment to any of the two conformers, they assigned the 866 cm$^{-1}$ band to the \ce{O-H-O} asymmetric stretch, the 982 cm$^{-1}$ band to the \ce{C-O} stretch, the features in the 1070-1225 cm$^{-1}$ region to a combination of the \ce{C-O-H} bend/\ce{CH3} rock/\ce{O-H-O} asymmetric stretch and the \ce{CH3} d-deformation/\ce{CH3} rock, the features in the 1330-1425 cm$^{-1}$ region to combinations of the in phase \ce{C-O-H} bend/\ce{CH3} rock/ \ce{O-H-O} asymmetric stretch, and the features in the 1467-1634 cm$^{-1}$ region to a combination of the \ce{CH3} s-deformation and the \ce{H-O-H} bend.

Overall, these observations are in good agreement with our observed bands at 851(2), 981(3), 1098(3), 1386(3) and 1553(3) cm$^{-1}$, although the three bands observed below 800 cm$^{-1}$ as part of this work were not reproduced by this earlier study. However, a slight unresolved shoulder to the band at 866 cm$^{-1}$ that is not assigned in the earlier work is a good qualitative match for the band we observe at 734(3) cm$^{-1}$. 

Given the high level of agreement between our experimental spectrum presented in Fig. \ref{fig:m65IRMPD} and that recorded previously 
\cite{Fridgen}, we conclude that the majority of the \textit{m/z} 65 ion signal observed for the reaction in the trap corresponds to a mixture of the two conformers of the methanol proton-bound dimer. However, as we observe additional minor bands below 800 cm$^{-1}$ as well as slight changes in the relative intensities of some bands from those of the two conformers of the methanol proton-bound dimer, we cannot disregard the possibility that other isomers are also present with low abundances. 

\begin{figure}
    \centering
    \includegraphics[width=13cm]{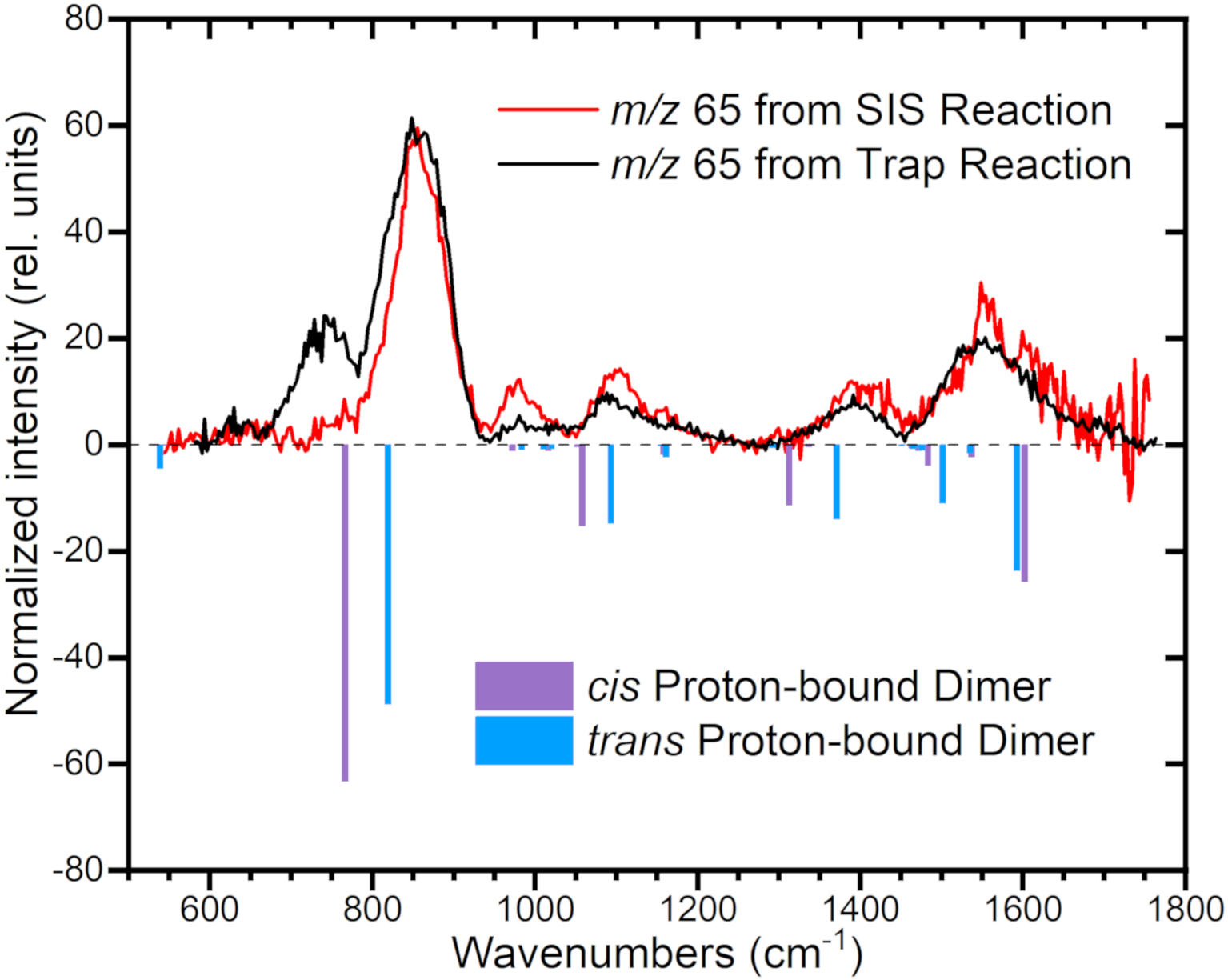}
    \caption{IR-MPD spectra of the \textit{m/z} 65 complexes formed by the reaction of \ce{CH3OH2+} with \ce{CH3OH} both in the trap (black) and in the SIS (red). Comparison with the calculated vibrational modes of \textit{cis} and \textit{trans} methanol proton-bound structures (purple and blue, respectively). Both experimental curves are normalized for the laser power to have a better comparison, and re-normalized for the strongest band intensity.}
    \label{fig:m65IRMPD_SIS}
\end{figure}

As mentioned earlier, as well as collecting an IR-MPD spectrum for the depletion of the \textit{m/z} 65 ion formed in the trap, we have also collected an equivalent IR-MPD spectrum of the \textit{m/z} 65 ion formed in the SIS, with both spectra, re-normalized using the intense band at $\sim 855$ cm$^{-1}$, shown together in Fig. \ref{fig:m65IRMPD_SIS}, along with the computed spectra for the two conformers of the methanol proton-bound dimer. 


From this, we note that the main band at 851(2) cm$^{-1}$ is well reproduced in the SIS spectrum, as are the higher wavenumber bands (though the intensities relative to the main band vary slightly). However, the band is much narrower, with the weaker feature at 734(3) cm$^{-1}$ being almost completely absent. In addition, the  minor bands at 624(3) and 649(3) cm$^{-1}$ are not present in the spectrum of \textit{m/z} 65 produced in the SIS.

Considering the spectra of the \textit{m/z} 65 intermediates formed in both the trap and the SIS we note three significant observations. Firstly, the most intense bands assigned to the \textit{cis} and \textit{trans} conformers have quite different intensities in the spectra of the ions formed in the trap. Secondly, this difference is even more pronounced for the spectra of the ions formed in the SIS, where the band assigned to the \textit{cis} conformer is almost completely absent. Thirdly, there is little to no evidence of species other than the proton-bound conformers in either spectra.

As the two conformers of the proton-bound dimer are connected via a transition state, \textbf{TS1}, the relative populations of the two can be approximated by the appropriate Boltzmann factors. For the ions formed in the trap, reactive collisions occur in a \ce{He} buffer kept at $\sim 200$ K in order to prevent methanol freeze-out. At this internal temperature, the populations of the \textit{trans} and \textit{cis} conformers are predicted to be in a 1.8 to 1  ratio, somehow consistent with that observed in Fig. \ref{fig:m65IRMPD}. 

For the ions formed in the SIS, although the temperature in the SIS is higher than that in the trap, once entering the trap the ions are cooled in a \ce{He} buffer kept at $\sim 30$ K, leading to a much lower internal temperature. Assuming an internal temperature of 30 K, the population of the \textit{trans} conformer becomes approximately 55 times higher than that of the \textit{cis} conformer.
Although we are unable to determine the exact internal temperature of the ions in both cases, the estimated temperature is in reasonable  agreement with the observed relative abundances of the two conformers and provides a rationalisation of the observed differences, both between the abundances of the two conformers, and the spectra collected for the two experimental conditions. 


The third observation, the dominant formation of the proton-bound complex, is relatively easily rationalised in light of the PES shown in Fig. \ref{fig:PES}. The \ce{CH3}-bound dimer is expected to be comparatively short-lived as both of the connecting transition states, \textbf{TS2} and \textbf{TS3}, are comparatively low-lying in energy. Similarly, the \ce{H2O}-release complex is either able to quickly isomerize via \textbf{TS4} or separate into the products. Finally, while the proton-bound complex of the products is much lower in energy, the resultant high level of internal energy upon formation, combined with the lack of a barrier to separation into the products, is also expected to make this species comparatively short-lived.

Finally, in a previous work \cite{Fridgen}, spectra were also collected for the gain in both \textit{m/z} 33 and 47 ion signal due to the resonant photo-fragmentation of the \textit{m/z} 65 complex. However, while we have been able to perform equivalent measurements for the \textit{m/z} 65 ion formed in the SIS, 
this has not been possible for the ion formed in the trap due to the overlap of the fragment ion signal with the reactant \textit{m/z} 33 and product \textit{m/z} 47 ions also present in the trap. For the sake of completeness, results of these measurements are shown in Fig. S8 in the ESI, but we are unable to draw any further conclusions than those made in literature \cite{Fridgen}, other than to note the high level of agreement between the results obtained here and in that work. 

\section{Conclusions}

This work provides the first spectroscopic confirmation that the \textit{m/z} 47 product formed by the reaction of protonated methanol (\ce{CH3OH2+}) with methanol (\ce{CH3OH}) is indeed protonated DME, \ce{(CH3)2OH+}. This has been achieved through the collection and assignment of high-resolution IR-PD spectra of \ce{Ne}-tagged ions in combination with quantum chemical calculations. It should also be noted that these measurements mark the first spectroscopic characterization of protonated DME, thereby providing reference data for its astronomical search with infrared observatories, and enabling a solid basis for  follow-up spectroscopic investigations at higher resolution.

Furthermore, we have recorded IR-MPD spectra of the \textit{m/z} 65 complexes formed by collisional stabilisation of the reaction intermediates. These have been assigned through comparison with quantum chemical calculations and previous experimental work \cite{Fridgen} to the proton-bound methanol dimers. For the reaction performed in the  22-pole trap at 200 K, we observe both the \textit{cis} and \textit{trans} proton-bound dimers of methanol. For the reaction performed in the higher-temperature (400 K) SIS, only the \textit{trans} conformer is observed due to the lower-temperature equilibrium achieved following collisions in the ion trap (held at $\sim$ 30 K) prior to spectroscopic probing. The dominance of the proton-bound dimers over other isomeric structures has been rationalised in light of the reaction PES and the expected short lifetimes of the other reaction intermediates. 

We conclude by highlighting some implication of the present work for astrochemistry: 

\textit{i)} The abundance of DME at different phases of the star formation processes and its substantial proton affinity ($\sim 792$ kJ/mol \cite{linstorm1998nist}, larger than not only that of CO but also of other iCOMs such as formaldehyde and methanol) suggest that its protonated variant might also be present, not only as a precursor in the formation of neutral DME as discussed here, but also as a result of efficient proton transfer reactions from the abundant interstellar ions \ce{H3+} and \ce{HCO+}.

\textit{ii)} The formation of interstellar DME via the gas phase reaction of protonated and neutral methanol implies the formation of the methanol proton-bound dimer. Although no clear-cut astronomical detection has been reported so far, strongly bound molecular aggregates involving ions, such as the methanol proton bound dimer that was observed in our experiment, have been considered as potential contributors to interstellar chemistry, as it has been suggested that they form readily in dense and low-temperature environments, where they may act as intermediates for the synthesis of complex interstellar molecules \cite{Vaida2006, Potapov2017}. 

It would therefore be interesting to extend the spectroscopic study of protonated DME as well as of its gas phase precursor, the methanol proton bound dimer, to the millimetre/submillimetre-wave region to provide rotational laboratory spectra to guide and support their radio-astronomical observations.   

\section{Acknowledgments}

We would like to dedicate this work, which combines ion reactivity with spectroscopic probing, to the memory of our mentor and colleague Dieter Gerlich, who was a pioneer in the study of ion-molecule reactivity and spectroscopy. Dieter's innovative developments of numerous versatile techniques to guide, trap and cryogenically cool ions have been and will be an inspiration to our research. 

We gratefully acknowledge the support of Radboud University and of NWO for providing the required beam time at the FELIX Laboratory, and the skillful assistance of the FELIX staff. We thank the Cologne Laboratory Astrophysics group for providing the FELion ion trap instrument for the current experiments and the Cologne Center for Terahertz Spectroscopy funded by the Deutsche Forschungsgemeinschaft (DFG, grant SCHL 341/15–1) for supporting its operation. V.R. also acknowledges funding for a PhD fellowship from the Dept. Physics, University of Trento.

\section*{Disclosure statement}

The authors have no conflicts of interest to declare. 

\section*{Funding}

The research leading to these results has received funding from LASERLAB-EUROPE (grant agreement no. 871124, European Union’s Horizon 2020 research and innovation programme), from the Deutsche Forschungsgemeinschaft (DFG, grant SCHL 341/15–1), and from NWO Exact and Natural Sciences through the use of supercomputer facilities at SURFsara in Amsterdam (NWO Rekentijd grant 2021.055). This project has received funding from the European Union’s Horizon 2020 research and innovation programme under the Marie Sklodowska Curie grant agreement No 811312 for the project ”Astro-Chemical Origins” (ACO) and from  MUR PRIN 2020 project n. 2020AFB3FX  "Astrochemistry beyond the second period elements".



\bibliographystyle{tfo}
\bibliography{BIB}

\end{document}